\documentclass[oldversion]{aa}
\usepackage{epsfig}
\usepackage{natbib}
\usepackage{lscape}
\usepackage{wasysym}
\bibpunct{(}{)}{;}{a}{}{,}
\begin{document}

\title{Metallicities for 6 nearby open clusters from high-resolution spectra of giant stars\thanks{Based on 
observations collected at the La Silla Parana Observatory,
ESO (Chile) with the UVES spectrograph at the 8.2-m Kueyen telescope, under
program 383.C-0170.}}

\subtitle{[Fe/H] values for a planet search sample}

\author{
  N.C. Santos\inst{1,2} \and
  C. Lovis\inst{3} \and	
  J. Melendez\inst{4} \and
  M. Montalto\inst{1} \and
  D. Naef\inst{3} \and
  G. Pace\inst{1}
  }

\institute{
    Centro de Astrof{\'\i}sica, Universidade do Porto, Rua das Estrelas, 4150-762 Porto, Portugal
    \and
    Departamento de F{\'\i}sica e Astronomia, Faculdade de Ci\^encias, Universidade do Porto, Portugal
    \and
    Observatoire de Gen\`eve, Universit\'e de Gen\`eve, 51 ch. des Maillettes, 1290 Sauverny, Switzerland
    \and
    Departamento de Astronomia do IAG/USP, Universidade de S\~ao Paulo, Rua do Mat\~ao 1226, Cidade Universit\'aria, 05508-900 S\~ao Paulo, SP, Brasil
}

\date{Received XXX; accepted XXX}

\abstract{
We present a study of the stellar parameters and iron abundances of 18 giant stars in 6 open clusters. The analysis was based on high-resolution
and high-S/N spectra obtained with the UVES spectrograph (VLT-UT2). The results complement {our previous study} 
where 13 clusters were already analyzed. The total sample of 18 clusters is part of a program to search for planets around giant stars. The results show that the 18 clusters 
cover a metallicity range between $-$0.23 and +0.23\,dex. Together with the derivation of the stellar masses, these metallicities will allow the metallicity and
mass effects to be disentangled when analyzing the frequency of planets as a function of these stellar parameters.
  \keywords{planetary systems: formation --
  	    Stars: abundances --
	    Stars: fundamental parameters --
	    Techniques: spectroscopic --
	    open clusters and associations: general 
	    }}

\authorrunning{Santos et al.}
\titlerunning{[Fe/H] abundances in 6 open clusters}
\maketitle

\section{Introduction}

An increasing amount of evidence exists that stellar mass is a key parameter regulating giant planet formation. 
For instance, the frequency of giant planets orbiting (low-mass) M-dwarfs is considerably lower than the
one found for FGK dwarfs \citep[][]{Bonfils-2005b,Endl-2006}. 
Higher mass stars also seem to have a higher frequency of orbiting planets \citep[][]{Lovis-2007,Johnson-2007}. 
This result is expected from the models of planetary formation 
\citep[][]{Laughlin-2004,Ida-2005,Kennedy-2008}, although a consensus does not exist on
this point \citep[][]{Kornet-2005,Boss-2006}.

Addressing the frequency of planets around stars of different masses is, however, not a simple
task. FGK dwarfs only occupy a narrow range in mass (roughly from 0.8 to 1.2 M$_ {\odot}$), making any study of the planet frequency-stellar
mass correlation difficult. For their lower mass counterparts, M-dwarfs, the radial-velocity method is often made difficult by the high stellar activity levels \citep[e.g.][]{Forveille-2009},
though this is, however, balanced by the higher amplitude signals expected. On the upper mass side main-sequence objects present a tougher case, because of the lack of spectral information (lines), together with the usually high rotation velocities of higher mass dwarfs. Although a few results
exist \citep[e.g.][]{Galland-2005}, the search for planets orbiting main sequence, intermediate-mass stars has not seen much success.

One way to circumvent this problem is to search for planets around intermediate-mass evolved (giant or subgiant) stars. 
A few giant planet candidates have indeed been announced around these kind of objects 
\citep[e.g.][]{Frink-2002,Sato-2003,Setiawan-2005,Hatzes-2006,Niedzielski-2007,Johnson-2007b}.
The problem with this approach is that it is very difficult to derive precise and uniform mass values for field giants \citep[see e.g.][]{Lloyd-2011}.

The key to solving this problem may be to search for planets orbiting clump giant stars in galactic open clusters. The position of the turnoff stars in the
cluster allows setting strong constraints on their masses, giving immediate information about the mass of the clump giants. Furthermore, observing clusters with different ages 
(and metallicities) will allow stars to be covered within a whole range of masses. From the different surveys for such objects, three giant planets have been detected so far, 
namely those orbiting \object{NGC2423No3}, \object{NGC4349No127} \citep[][]{Lovis-2007}, and \object{$\epsilon$\,Tau} \citep[][]{Sato-2007}.

In \citet[][hereafter Paper I]{Santos-2009} we derived atmospheric parameters and chemical abundances
for giant and dwarfs stars in 13 clusters that are being monitored 
with radial velocities to search for giant planets \citep[][]{Lovis-2007}. In the present
paper we complement this study by deriving precise stellar parameters and iron abundances
in a sample of giant stars in five more clusters that are part of the same planet search program (plus one cluster
in common with Paper I).
In Sect.\,\ref{sec:sample} we present our sample and the observations. In Sect\,\ref{sec:analysis} we present the
analysis of the data. We conclude with a short discussion in Sect.\,\ref{sec:discussion}.

\section{Sample and data}
\label{sec:sample}

In Paper I we presented the stellar parameters and chemical abundances for 39 giants and 16 dwarfs in the 13 open clusters
\object{IC2714}, 
\object{IC4651},  
\object{IC4756},  
\object{NGC2360}, 
\object{NGC2423},
\object{NGC2447} (M93), 
\object{NGC2539}, 
\object{NGC2682} (M67), 
\object{NGC3114}, 
\object{NGC3680}, 
\object{NGC4349}, 
\object{NGC5822}, and
\object{NGC6633}.
These 13 clusters are part of the 18 being surveyed for planets by \citet[][]{Lovis-2007}. To complement the study
presented in Paper I, we present here a detailed spectroscopic analysis of 15 giants in the five remaining clusters, namely
\object{NGC2287},
\object{NGC2567},
\object{NGC3532},
\object{NGC6494}, and
\object{NGC6705}.
As a consistency check, three giants in NGC2447 were also observed again, even though this
cluster had already been analyzed in Paper I (using a different data set).

For each cluster, three giants were chosen based on CORALIE\footnote{At the 1.2-m Euler Swiss Telescope, La Silla, Chile} and HARPS\footnote{At the 3.6-m ESO telescope, La Silla, Chile} 
data collected as part of the planet search program \citep[for details we refer to][]{Lovis-2007}. CORALIE and HARPS data allow us to assure that the giants are cluster members (their radial-velocities are compatible with
membership), as well as that they are not short period binary stellar systems (e.g. spectroscopic binaries). 

High-resolution UVES spectra \citep[at the VLT-UT2 Kueyen telescope --][]{Dekker-2000} were obtained for each of the 18 giants. 
The observations were carried out in service mode in March 2009, under ESO program 383.C-0170.
As for the data presented in Paper I, the spectra were taken in the RED580 mode. The resulting spectra 
cover the wavelength domain between 4780 and 6805\AA, with 
a gap between 5730 and 5835\AA\ (corresponding to the gap in the
CCD mosaic). A slit width of 0.3 arcsec was adopted, providing a resolution
$\mathrm{R=\lambda/\Delta\lambda\approx100\,000}$. This is significantly higher than the one presented
in Paper\,I ($\mathrm{R=50\,000}$). The data for NGC2447, available in both formats, allows us to verify if using
a different resolution will have any systematic effect on the derived stellar parameters and
iron abundances. The S/N of the final spectra
is around 200 for all stars.

\begin{table}[t]
\caption{Target list of UVES-VLT program 383.C-0170. Data from \citet[][]{Mermilliod-2003}.}
\label{tab:target}
\begin{tabular}{lllr}
\hline\hline
\noalign{\smallskip}
Star  & $\alpha$ (2000.0) & $\delta$ (2000.0) & V \\
\hline
\multicolumn{4}{l}{NGC2287}\\
No21	& 06:45:57.46 & $-$20:46:30.2 & 6.91    \\      
No75	& 06:45:43.02 & $-$20:51:09.6 & 7.50    \\ 
No97	& 06:46:04.84 & $-$20:36:24.9 & 7.78    \\ 
\multicolumn{4}{l}{NGC2447}\\
No28   & 07:44:50.25 & $-$23:52:27.1 & 9.85     \\  
No34   & 07:44:33.66 & $-$23:51:42.2 & 10.12    \\  
No41   & 07:44:25.73 & $-$23:49:53.0 & 10.03    \\ 
\multicolumn{4}{l}{NGC2567}\\
No16   & 08:18:35.26 & $-$30:38:57.9 & 11.04    \\ 
No54   & 08:18:26.43 & $-$30:39:30.4 & 11.16    \\ 
No114 & 08:18:19.18 & $-$30:32:58.4 & 10.87    \\ 
\multicolumn{4}{l}{NGC3532}\\
No19   & 11:05:58.74 & $-$58:43:29.4 & 7.74    \\ 
No100 & 11:06:03.84 & $-$58:41:15.8 & 7.50    \\ 
No122 & 11:05:45.63 & $-$58:40:39.4 & 8.20    \\ 
\multicolumn{4}{l}{NGC6494}\\
No6   & 17:56:51.99 & $-$19:00:03.4 & 9.68    \\ 
No48 & 17:56:23.04 & $-$19:08:58.7 & 9.57    \\ 
No49 & 17:56:41.17 & $-$19:08:38.3 & 9.70    \\ 
\multicolumn{4}{l}{NGC6705}\\
No1090 & 18:51:03.99 & $-$06:20:41.0 & 11.87    \\ 
No1184 & 18:51:02.02 & $-$06:17:26.2 & 11.43    \\ 
No1111 & 18:51:03.60 & $-$06:16:11.0 & 11.90    \\ 
\hline
\noalign{\smallskip}
\end{tabular}
\end{table}

\begin{table*}
\caption[]{Stellar parameters derived for the giants stars. See text for more details.}
\begin{tabular}{lcccrcc}
\hline
Star     & $T_{\mathrm{eff}}$ & $\log{g}_\mathrm{spec}$ & $\xi_{\mathrm{t}}$ & [Fe/H] & N(\ion{Fe}{i},\ion{Fe}{ii}) & $\sigma$(\ion{Fe}{i},\ion{Fe}{ii})  \\
            &   [K] &  [g in cm\,s$^{-2}$]  &  [km\,s$^{-1}$]  &  &  &   \\
\hline
\multicolumn{7}{l}{Results using the S08 line-list:}\\
NGC2287No21$\dagger$  &  4350$\pm$82  &  1.71$\pm$0.66  &  2.49$\pm$0.08  &  -0.24$\pm$0.24  &  179,25  &  0.24,0.37\\
NGC2287No75  &  4617$\pm$57  &  2.03$\pm$0.37  &  2.20$\pm$0.05  &  $-$0.12$\pm$0.18  &  179,24  &  0.18,0.20\\
NGC2287No97  &  4764$\pm$46  &  2.15$\pm$0.40  &  1.98$\pm$0.04  &  $-$0.09$\pm$0.14  &  181,25  &  0.14,0.20\\
NGC2447No28  &  5143$\pm$28  &  2.76$\pm$0.26  &  1.61$\pm$0.03  &  $-$0.05$\pm$0.09  &  180,25  &  0.09,0.12\\
NGC2447No34  &  5242$\pm$35  &  3.01$\pm$0.29  &  1.63$\pm$0.03  &  0.03$\pm$0.11  &  179,25  &  0.11,0.13\\
NGC2447No41  &  5215$\pm$25  &  2.94$\pm$0.36  &  1.59$\pm$0.02  &  0.00$\pm$0.08  &  179,25  &  0.08,0.16\\
NGC2567No16  &  5205$\pm$42  &  2.82$\pm$0.36  &  1.59$\pm$0.04  &  $-$0.04$\pm$0.14  &  182,23  &  0.13,0.16\\
NGC2567No54  &  5216$\pm$36  &  2.89$\pm$0.34  &  1.73$\pm$0.04  &  0.06$\pm$0.12  &  182,23  &  0.12,0.16\\
NGC2567No114  &  5078$\pm$39  &  2.73$\pm$0.33  &  1.89$\pm$0.04  &  0.01$\pm$0.14  &  180,24  &  0.13,0.16\\
NGC3532No19  &  5089$\pm$34  &  2.69$\pm$0.22  &  1.66$\pm$0.03  &  0.05$\pm$0.11  &  182,25  &  0.11,0.10\\
NGC3532No100  &  4938$\pm$41  &  2.54$\pm$0.21  &  1.85$\pm$0.04  &  0.00$\pm$0.13  &  179,24  &  0.13,0.10\\
NGC3532No122  &  5218$\pm$55  &  3.09$\pm$0.71  &  1.74$\pm$0.06  &  0.01$\pm$0.17  &  178,25  &  0.16,0.31\\
NGC6494No6  &  4926$\pm$42  &  2.50$\pm$0.24  &  1.90$\pm$0.04  &  0.02$\pm$0.14  &  181,24  &  0.13,0.12\\
NGC6494No48  &  5131$\pm$46  &  2.71$\pm$0.46  &  1.86$\pm$0.05  &  0.04$\pm$0.16  &  179,24  &  0.15,0.21\\
NGC6494No49  &  5012$\pm$48  &  2.63$\pm$0.36  &  1.82$\pm$0.04  &  0.06$\pm$0.16  &  181,25  &  0.15,0.17\\
NGC6705No1090  &  4844$\pm$70  &  2.43$\pm$0.6  &  2.06$\pm$0.06  &  0.12$\pm$0.22  &  181,24  &  0.21,0.31\\
NGC6705No1111  &  5039$\pm$74  &  2.85$\pm$0.54  &  2.18$\pm$0.08  &  0.14$\pm$0.24  &  184,24  &  0.23,0.27\\
NGC6705No1184$\dagger$  &  4518$\pm$65  &  2.09$\pm$0.42  &  1.92$\pm$0.06  &  $-$0.01$\pm$0.19  &  178,23  &  0.19,0.23\\
\hline
\multicolumn{7}{l}{Results using the HM07 line-list:}\\
NGC2287No75$\dagger$  &  4381$\pm$135  &  1.73$\pm$0.33  &  2.11$\pm$0.15  &  $-$0.20$\pm$0.17  &  16,6  &  0.15,0.07\\
NGC2287No97  &  4638$\pm$84  &  2.14$\pm$0.20  &  1.87$\pm$0.10  & $-$0.11$\pm$0.11  &  15,6  &  0.09,0.07\\
NGC2447No28  &  5077$\pm$97  &  2.90$\pm$0.14  &  1.71$\pm$0.18  &  $-$0.08$\pm$0.14  &  15,6  &  0.10,0.03\\
NGC2447No34  &  5064$\pm$95  &  2.90$\pm$0.17  &  1.76$\pm$0.22  &  $-$0.10$\pm$0.16  &  16,6  &  0.12,0.06\\
NGC2447No41  &  5109$\pm$75  &  2.72$\pm$0.40  &  1.60$\pm$0.15  &  $-$0.07$\pm$0.12  &  16,6  &  0.09,0.19\\
NGC2567No16  &  5061$\pm$85  &  2.72$\pm$0.24  &  1.67$\pm$0.19  &  $-$0.17$\pm$0.14  &  16,6  &  0.10,0.11\\
NGC2567No54  &  5040$\pm$93  &  2.66$\pm$0.30  &  1.73$\pm$0.16  &  $-$0.04$\pm$0.15  &  16,6  &  0.11,0.14\\
NGC2567No114  &  4890$\pm$98  &  2.59$\pm$0.21  &  1.86$\pm$0.14  &  $-$0.09$\pm$0.14  &  16,6  &  0.10,0.08\\
NGC3532No19  &  4944$\pm$66  &  2.47$\pm$0.19  &  1.57$\pm$0.10  &  0.00$\pm$0.11  &  16,6  &  0.08,0.09\\
NGC3532No100  &  4766$\pm$74  &  2.22$\pm$0.22  &  1.71$\pm$0.09  &  $-$0.07$\pm$0.12  &  15,6  &  0.09,0.10\\
NGC3532No122  &  4929$\pm$81  &  2.51$\pm$0.51  &  1.41$\pm$0.14  &  $-$0.14$\pm$0.13  &  15,6  &  0.10,0.25\\
NGC6494No6  &  4779$\pm$76  &  2.27$\pm$0.15  &  1.78$\pm$0.10  &  $-$0.07$\pm$0.11  &  15,6  &  0.08,0.04\\
NGC6494No48  &  4909$\pm$84  &  2.54$\pm$0.18  &  1.90$\pm$0.17  &  $-$0.14$\pm$0.13  &  16,6  &  0.10,0.07\\
NGC6494No49  &  4795$\pm$107  &  2.26$\pm$0.32  &  1.80$\pm$0.14  &  $-$0.11$\pm$0.15  &  16,6  &  0.11,0.14\\
NGC6705No1111  &  4685$\pm$190  &  2.28$\pm$0.52  &  2.26$\pm$0.28  &  $-$0.06$\pm$0.25  &  16,6  &  0.19,0.22\\
NGC6705No1184$\dagger$  &  4288$\pm$88  &  1.72$\pm$0.31  &  1.96$\pm$0.09  &  $-$0.14$\pm$0.09  &  15,6  &  0.08,0.13\\
\hline
\end{tabular}\\
$\dagger$ Given their low derived effective temperatures, these measurements were excluded when computing the metallicities listed in Table\,\ref{tab:average}.
\label{tab:giants}
\end{table*}

\section{Stellar parameters and iron abundances}
\label{sec:analysis}

Stellar atmospheric parameters and iron abundances were derived 
in LTE using the 2002 version of the code 
MOOG \citep[][]{Sneden-1973}\footnote{http://verdi.as.utexas.edu/moog.html} 
and a grid of Kurucz Atlas plane-parallel model atmospheres \citep[][]{Kurucz-1993}. 
Parameters were obtained by imposing excitation and ionization 
equilibrium to a set of \ion{Fe}{i} and \ion{Fe}{ii} lines, following
the basic prescription described in \citet[][]{Santos-2004b}.
In brief, the parameters were derived through an iterative process until the slope of the relation between
the abundances given by individual \ion{Fe}{i} lines and both the excitation 
potential ($\chi_l$) and reduced equivalent width ($\log{EW/\lambda}$) were zero,
and until the FeI and FeII lines provided the same average abundance.
The adopted solar abundances are from \citet[][]{Anders-1989} except for iron, for which we adopt a 
value of $\log{\epsilon \mathrm{(Fe)}}$=7.47 \citep[the same as used in Paper\,I, taken from][]{Gonzalez-2000}.
Given that we are doing a differential analysis relative to the Sun, we do not expect that our results have any significant dependence on the adopted solar abundances. 
Individual line equivalent widths (EW) for the iron lines were measured using the
automatic ARES code \citep[][]{Sousa-2007,Sousa-2008}\footnote{http://www.astro.up.pt/$\sim$sousasag/ares/}.

As in Paper I, the stellar parameters and iron abundances for our giant stars were derived 
using two different line lists. First, we used the line list described 
in \citet[][hereafter S08]{Sousa-2008}, composed of 263 \ion{Fe}{i} and 36 \ion{Fe}{ii} lines
in the optical domain. The use of this line list has shown to give excellent results for the analysis 
of dwarf stars \citep[][]{Sousa-2008}. Beyond this, we also derived the stellar
parameters using the line list provided by \citet[][hereafter HM07]{Hekker-2007}. Although
much shorter (20 \ion{Fe}{i} and 6 \ion{Fe}{ii} lines), the lines
in this list were carefully chosen for the analysis of giant stars, avoiding line-blending
from CN lines \citep[][]{Melendez-1999}.

\begin{figure*}[t!]
\resizebox{8.8cm}{!}{\includegraphics{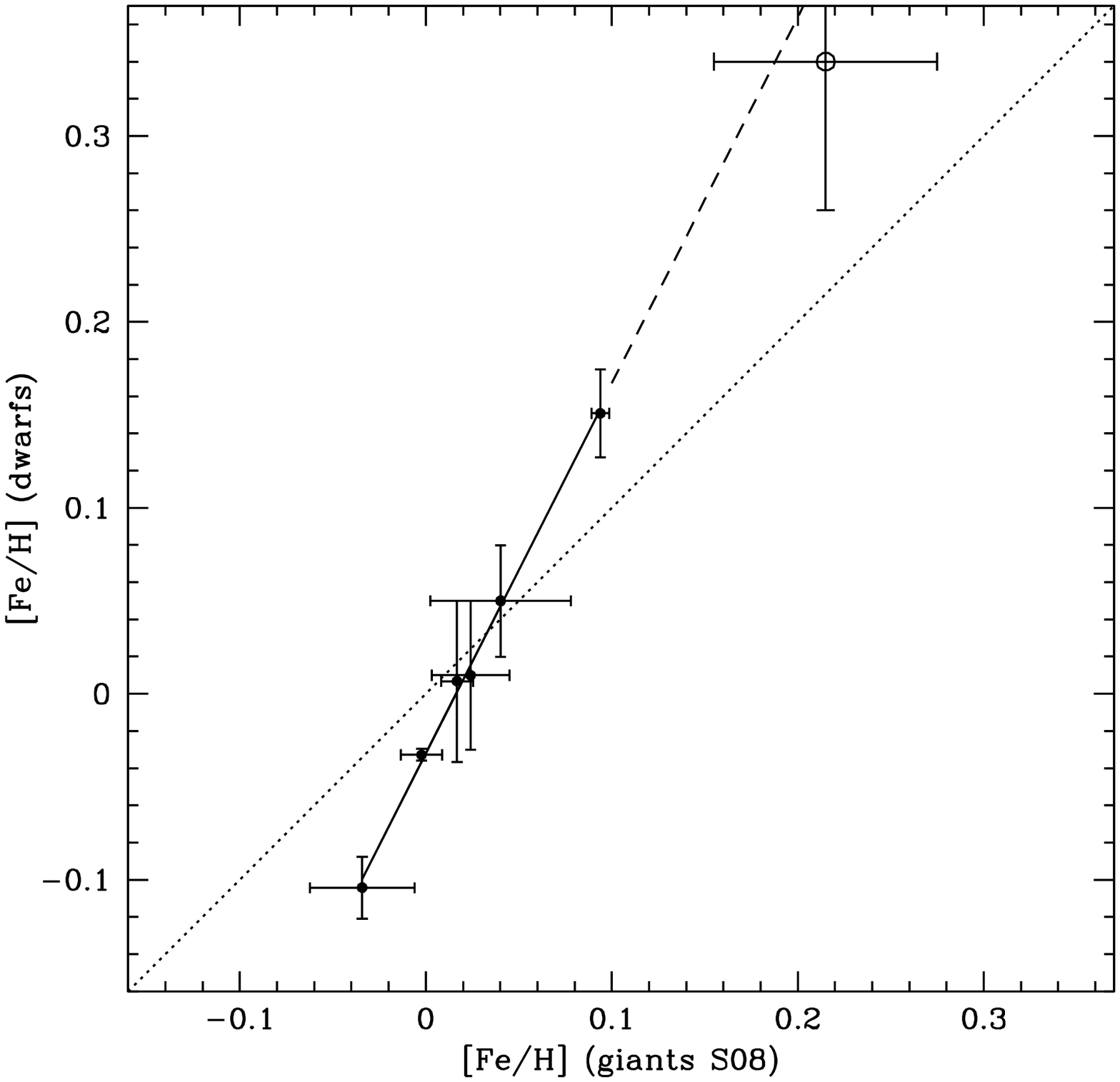}}
\resizebox{8.8cm}{!}{\includegraphics{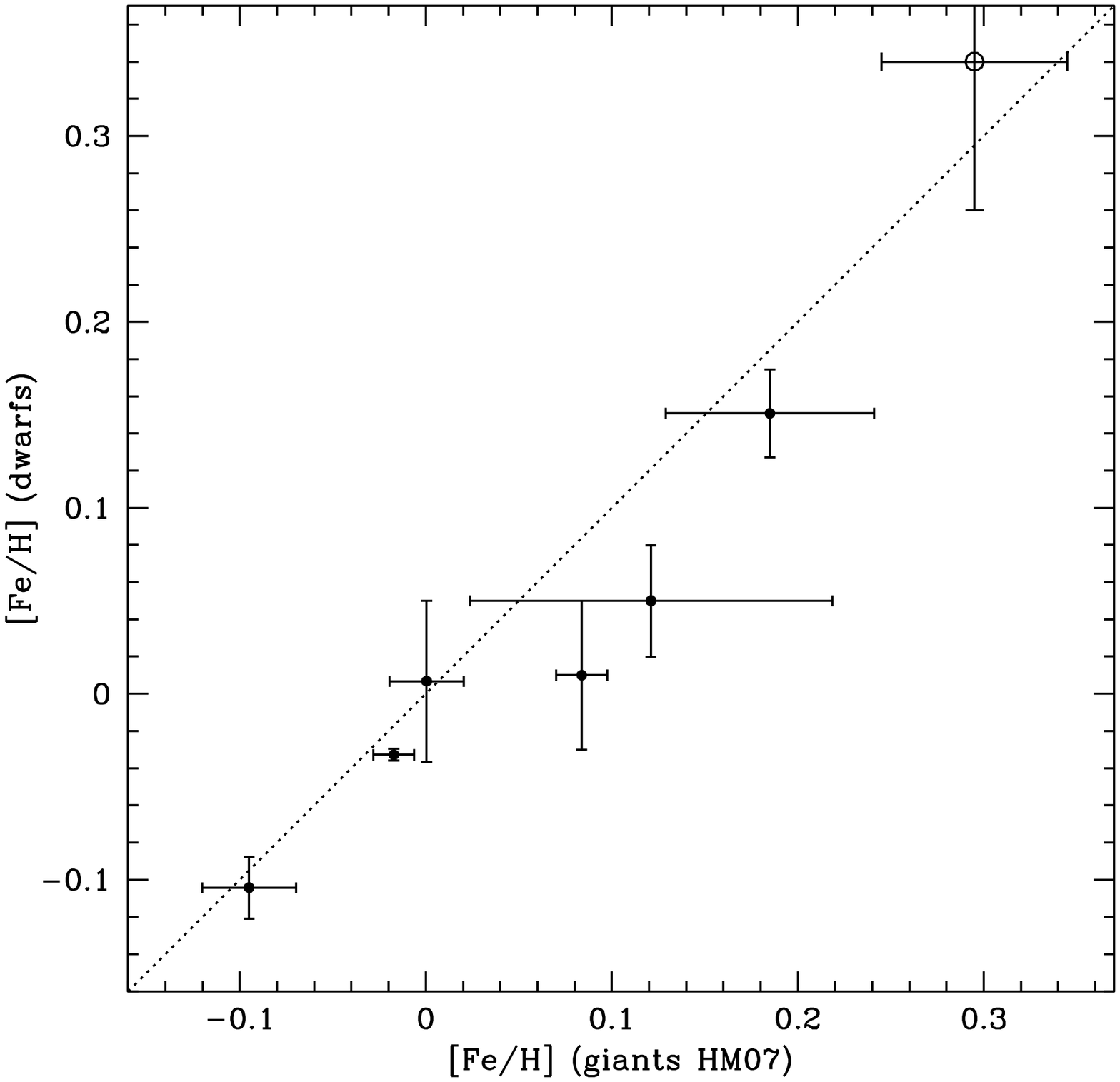}}
\caption{Comparison of the average values for the metallicities derived from
dwarfs and giants in the 6 clusters from Paper I (dots), as well as for dwarfs and giants from
NGC6253 \citep[][open circle]{Montalto-2012}. See text for more details.}
\label{fig:comparison}
\end{figure*}

As shown in Paper I, the parameters derived using the HM07 line list may be
preferable since they provide metallicities on the same scale as the ones derived for dwarf stars.
However, since they are more subject to analysis errors (much shorter line list), 
in Paper I we proposed a correction for the [Fe/H] values derived using the S08 line-list.
This correction puts the derived metallicities for the giant stars on the same scale as those found for dwarfs of the
same cluster. For that we used the average metallicities derived from dwarfs and
giants in six different clusters, spanning a metallicity range roughly between $-$0.1
and $+$0.2\,dex (see also Fig.\,\ref{fig:comparison}, left panel).

The derived (uncorrected) stellar parameters and iron abundances for all the stars analyzed
in this paper are listed in Table\,\ref{tab:giants}, together with
the number of \ion{Fe}{i} and \ion{Fe}{ii} lines used and the rms of their individual
abundances. The cluster-averaged metallicity
values can be seen in Table\,\ref{tab:average}, where we list the values
derived using both the S08 and the HM07 line lists. The corrected metallicities,
$<$[Fe/H]$>$$^\mathrm{c}_{\rm S08}$, refer to the values derived using the line list of S08 and subsequently
corrected using the relation presented in Paper I (illustrated in Fig.\,\ref{fig:comparison}, left panel -- see discussion in Sect.\,\ref{sec:comparison}). 
We adopt these metallicities as our final values.

For some of the stars in our sample, the small number of
lines in the HM07 line list did not allow us to derive reliable parameters (e.g. measurable lines
in the spectra did not cover a wide enough excitation potential range).
This was the case for stars No21 and No1090 in NGC2287 and NGC6705, respectively.

For some of the stars we derived effective temperature values below 4600\,K. These are the cases
of NGC2287No21 and NGC6705No1184 (using the S08 line list) and NGC2287No75 and NGC6705No1184 (using the HM07 line-list). 
Since we are not sure that the model atmospheres used are valid for such low temperatures {and considering that line
blending is more severe at lower temperatures}, we decided to keep these stars out
of the sample when computing the average abundances in these clusters (Table\,\ref{tab:average}). 
When one single star is left, the value in the table denotes the metallicity (and error)
derived for this star

Finally, we note that for NGC2287, the metallicity found using the S08 line list puts this cluster
outside the range for which the correction was derived (see Fig.\,\ref{fig:comparison} and Sect.\,\ref{sec:comparison}).
The obtained $<$[Fe/H]$>$$^\mathrm{c}_{\rm S08}$ value must in this case be taken with care. 
As described in Sect.\,\ref{sec:comparison}, while for high metallicities the results for NGC6253 seem to confirm that the calibration
is valid in the high-[Fe/H] regime, we have no confirmation for lower metallicity values.

\subsection{The metallicity scale for giants and dwarfs}
\label{sec:comparison}

In Fig.\,\ref{fig:comparison} we show two plots similar to those presented in Fig.\,1 of Paper I,
where we compare the metallicities derived for the dwarfs and giants in the six clusters
presented in that paper. Additionally, in the plot we include the metallicities derived for \object{NGC6253}, 
a metal rich cluster whose abundances were recently derived by our team and presented 
in \citet[][]{Montalto-2012}\footnote{The values derived are: 
for the only dwarf analyzed, [Fe/H]=+0.34$\pm$0.08; for the two giants analyzed, the average metallicities and 
their rms are +0.22$\pm$0.06 and +0.30$\pm$0.05, respectively using the S08 and HM07 line lists.}. 
In both panels, the dotted line represents the 1:1 relation, while the solid line in the left panel represents
a linear fit to the filled points, with a relation $\mathrm{<[Fe/H]>_{dwarfs}=1.97\,<[Fe/H]>_{S08}-0.03}$ (see Paper I).
The dashed line represents the extrapolation of this fit to higher metallicities. S08 and HM07 refer to the abundances derived
using the line lists of \citet[][]{Sousa-2008} and \citet[][]{Hekker-2007}, respectively.

\begin{table*}[t]
\caption[]{Weighted average metallicities of the giant stars in each of the 18 clusters (analysed in this paper and in Paper I). }
\begin{tabular}{lrrrl}
\hline
Cluster     & $<$[Fe/H]$>$$_{\mathrm{S08}}$ & $<$[Fe/H]$>$$^\mathrm{c}_{\mathrm{S08}}$ & $<$[Fe/H]$>$$_{\mathrm{HM07}}$ & Souce \\
\hline
IC2714 &  0.02$\pm$0.01 &  0.01$\pm$0.01 & $-$0.03$\pm$0.04& Paper I \\
IC4651 &  0.09$\pm$0.01 &  0.15$\pm$0.01 &  0.19$\pm$0.06& Paper I  \\
IC4756 &  0.02$\pm$0.02 &  0.02$\pm$0.02 &  0.08$\pm$0.01 & Paper I \\
NGC2360 &  0.00$\pm$0.01 & $-$0.03$\pm$0.01 & $-$0.01$\pm$0.03 & Paper I \\
NGC2423 &  0.09$\pm$0.06 &  0.14$\pm$0.06 &  0.07$\pm$0.06 & Paper I \\
NGC2447$^\dagger$ &$-$0.03$\pm$0.03 & $-$0.10$\pm$0.03 & $-$0.10$\pm$0.03 & Paper I \\
NGC2539 &  0.08$\pm$0.03 &  0.13$\pm$0.03 &  0.09$\pm$0.02 & Paper I \\
NGC2682 &  0.02$\pm$0.01 &  0.00$\pm$0.01 &  0.00$\pm$0.02 & Paper I \\
NGC3114 &  0.02$\pm$0.09 &  0.02$\pm$0.09 &  0.00$\pm$0.12 & Paper I \\
NGC3680 &  0.00$\pm$0.01 & $-$0.04$\pm$0.01 & $-$0.02$\pm$0.01 & Paper I \\
NGC4349 & $-$0.04$\pm$0.06 & $-$0.12$\pm$0.06 & $-$0.06$\pm$0.08 & Paper I \\
NGC5822 &  0.04$\pm$0.04 &  0.05$\pm$0.04 &  0.12$\pm$0.10 & Paper I \\
NGC6633 &  0.04$\pm$0.01 &  0.06$\pm$0.01 &  0.00$\pm$0.00 & Paper I \\
\hline
NGC2287 &  $-$0.10$\pm$0.02 &  $-$0.23$\pm$0.02 &  $-$0.11$\pm$0.11& This paper \\
NGC2447$^\dagger$ &  $-$0.01$\pm$0.04 &  $-$0.05$\pm$0.04 &  $-$0.08$\pm$0.01& This paper \\
NGC2567 &  0.01$\pm$0.05 &  0.00$\pm$0.05 &  $-$0.10$\pm$0.06& This paper \\
NGC3532 &  0.03$\pm$0.03 &  0.02$\pm$0.03 &  $-$0.06$\pm$0.07 & This paper \\
NGC6494 &  0.04$\pm$0.02 &  0.04$\pm$0.02 &  $-$0.10$\pm$0.04& This paper \\
NGC6705 &  0.13$\pm$0.01 &  0.23$\pm$0.01 &  $-$0.06$\pm$0.25& This paper \\
\hline
\end{tabular}\\
$^\dagger$ For this cluster we adopt the average value of the two derivations.
\label{tab:average}
\end{table*}

As can be seen from Fig.\,\ref{fig:comparison}, the position of NGC6253 seems to confirm our previous finding that the S08 line list,
despite its adequacy for the study of dwarf stars  \citep[][]{Sousa-2008}, does not deliver [Fe/H] values on the 
same scale for dwarfs and giant stars. On the other hand, the HM07 line list seems adequate for studying giants. The
calibration proposed in Paper I to correct the metallicities derived using the S08 line list also seems
to be appropriate. We decided not to change this calibration using the
new values for NGC6253 since only one dwarf has been analyzed in our study of this cluster \citep[][]{Montalto-2012}.

The cause of the offset metallicity values found using the S08 line list is not clear. In Paper I we suggested
that it could be because (cool) giant stars have higher
macroturbulence velocities \citep[][]{Gray-1992} and because thousands of molecular lines (CN, C2, CH, MgH) 
contribute to the optical spectra of giants \citep[][]{Coelho-2005}. These effects lead to line-blending,
which is stronger for higher metallicity 
and cooler objects. Indeed, in the present work the coolest giants ($<$4600 K)
seem more metal-poor than the hottest giants in the
same cluster (see results for NGC\,2287 and NGC\,6705), which could be explained by
severe line crowding, leading to underestimating the
true continuum and therefore to underestimating their metallicities.
Interestingly, the problem with the metallicity scale for giants 
(with respect to dwarfs) has also been recognized by other studies of giants in the solar 
neighborhood \citep[][]{Taylor-2005}, as well as of bulge stars \citep[e.g.][]{Cohen-2008}.

\begin{figure}[t!]
\resizebox{\hsize}{!}{\includegraphics{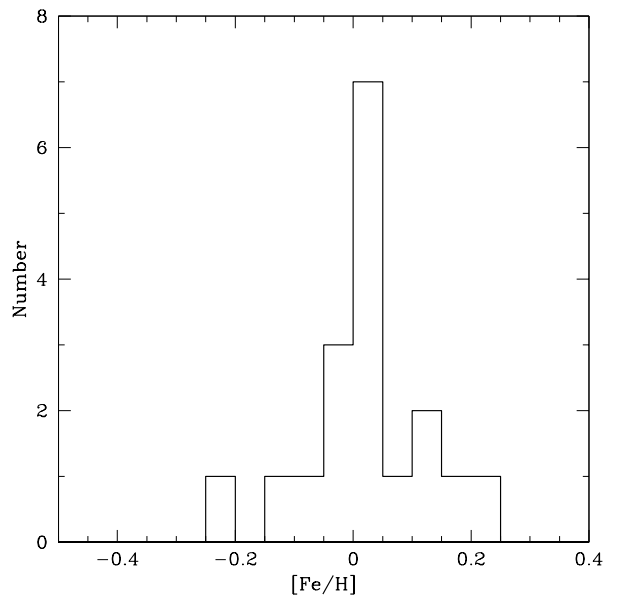}}
\caption{Metallicity distribution of the average metallicities for the
18 clusters in Table\,\ref{tab:average} using the corrected S08 values, $<$[Fe/H]$>$$^\mathrm{c}_{\rm S08}$.} 
\label{fig:histo}
\end{figure}

Using higher resolution spectra could in principle
minimize this problem. However, comparison of the results for NGC2447, whose spectra in Paper I were
taken with a resolution of $\approx$50\,000, while here we use data with a resolution of $\approx$100\,000 (for the
very same stars), does not confirm this possibility. Both sets of spectra provide very similar results in terms of
stellar metallicity (see the $<$[Fe/H]$>$$_{\mathrm{S08}}$ and $<$[Fe/H]$>$$_{\mathrm{HM07}}$ values in Table\,\ref{tab:average}), 
although the small observed difference
goes in the right direction: slightly higher metallicities
are obtained using spectra of higher resolution (even if the values are compatible within the errors).
On the other hand, this gives
us confidence that the results are homogeneous and can be treated together. For
the remainder of this paper we use the average metallicity values ($<$[Fe/H]$>$$^\mathrm{c}_{\mathrm{S08}}$=$-$0.07\,dex) derived in the two papers
for this cluster.

{
To test whether a wrong estimate of the surface gravities could be the cause of the observed discrepancy, 
we tested the influence of fixing $\log{g}$ to 2.50\,dex in NGC2447No41 i.e., $\approx$0.5\,dex below the
value obtained using the S08 line list. From this procedure we determined values of 5252\,K, 1.67\,km\,s$^{-1}$, and 0.00\,dex
for the derived effective temperature, microturbulence, and stellar metallicity respectively. 
The variations with respect to the value listed in Table\,\ref{tab:giants} are always within the error bars, and the derived metallicity did not suffer any change.
}

\section{Discussion}
\label{sec:discussion}

In Fig.\,\ref{fig:histo} we present the metallicity distribution for the 13 clusters studied in
Paper I, together with the five more clusters presented in this paper. Their metallicities (we are considering the  $<$[Fe/H]$>$$^\mathrm{c}_{\mathrm{S08}}$ values)
range from $-$0.23 to +0.23\,dex, with a peak near solar values. Most of the clusters seem, however, to
have metallicities above solar (only 6 among the 18 have [Fe/H]$<$0).

The knowledge about the metallicity of these clusters is of utmost importance for
interpretating the planet search results. It is well known that the metallicity of the star
is a prime factor in determining the frequency of giant planets \citep[][]{Santos-2001,Santos-2004b,Fischer-2005,Sousa-2011}.
Present results suggest that the observed metallicity-giant planet correlation reflects the higher
probability of forming planets orbiting metal-rich stars \citep[e.g.][]{Ida-2004b,Mordasini-2009a}.

This possibility has been put in question by the suggestion that the metallicity-giant planet correlation
may not hold for intermediate-mass (giant) stars hosting giant planets \citep[][]{Pasquini-2007,Ghezzi-2010}.
Although not fully accepted \citep[][]{Hekker-2007}, this possible lack
of correlation could hint that stellar mass strongly influences the planet formation process \citep[][]{Laughlin-2004,Ida-2005,Kennedy-2008}.
In either case, since both mass and metallicity seem to influence the planet formation process, the knowledge about the stellar
metallicity is crucial if one wants to disentangle both effects in the analysis of the results from the planet search program.

Since the number of stars surveyed in each cluster is not very high \citep[a dozen stars on average --][]{Lovis-2007}, 
knowledge of their metallicity may also constrain which clusters are more likely to harbor planets, and thus
focus the efforts of the survey. Alternatively, considering the metallicity-giant planet correlation observed for dwarfs,
it may be also wise to increase the sample in the less metal-rich clusters.

Until the present date, two giant planets have been announced as orbiting stars from the surveyed clusters: \object{NGC2423No3} and \object{NGC4349No127} \citep[][]{Lovis-2007}.
The two clusters have average metallicities of $+$0.14$\pm$0.06 and $-$0.12$\pm$0.06,
respectively (using the corrected values in Table\,\ref{tab:average}). Such a small number of planets preclude a statistical analysis of these results. 
However, it is curious to see that these two clusters are among the most metal-rich (\object{NGC2423}) and metal-poor (\object{NGC4349})
populations in the sample. 

\begin{acknowledgements}
This work was supported by the European Research Council/European Community under the FP7 through Starting Grant agreement 
number 239953. NCS also acknowledges the support from Funda\c{c}\~ao para a Ci\^encia e a Tecnologia (FCT) through program 
Ci\^encia\,2007 funded by FCT/MCTES (Portugal) and POPH/FSE (EC), and in the form of grant reference PTDC/CTE-AST/098528/2008.
MM also acknowledge the support from FCT in the form of fellowship reference SFRH/BDP/71230/2010. This research has made use of the WEBDA database, 
operated at the Institute for Astronomy of the University of Vienna.
\end{acknowledgements}

\bibliographystyle{aa}
\bibliography{santos_bibliography}

\end{document}